\documentclass[useAMS,usenatbib]{mn2e}
\usepackage{amsmath}
\usepackage{amssymb}
\usepackage{float}
\usepackage{graphicx}
\usepackage{lscape}
\usepackage{url}
\usepackage{natbib}
\usepackage[version=4]{mhchem}
\usepackage[utf8]{inputenc}

\newcommand{\pup}{$L_2$ Puppis }
\title[3-D hydrodynamic simulations of \pup]{Three-dimensional hydrodynamic simulations of \pup}

\author[Z. Chen et al.]
{Zhuo Chen,$^{1}$\thanks{E-mail: zchen25@ur.rochester.edu}
Jason Nordhaus,$^{2,3}$\thanks{E-mail: nordhaus@astro.rit.edu}
Adam Frank,$^{1}$
Eric ~G. Blackman$^{1}$\newauthor
and Bruce Balick$^{4}$\\
$^{1}$Department of Physics and Astronomy, University of Rochester, Rochester NY, 14627\\
$^{2}$Dept. of Science and Mathematics, National Technical Institute for the Deaf, Rochester Institute of Technology, Rochester, NY 14623\\
$^{3}$Center for Computational Relativity and Gravitation, Rochester Institute of Technology, Rochester, NY 14623\\
$^{4}$Department of Astronomy, University of Washington, Seattle, WA 98195, USA}
\begin{document}

\date{in original form 2016 February 19}

\pagerange{\pageref{firstpage}--\pageref{lastpage}} \pubyear{2016}

\maketitle

\label{firstpage}

\begin{abstract}
Recent observations of the \pup system suggest that this Mira-like variable may be in the early stages of forming a bipolar planetary nebula (PN).  As one of nearest and brightest AGB stars, thought be a binary, \pup serves as a benchmark object for studying the late-stages of stellar evolution.  We perform global, three-dimensional, adaptive-mesh-refinement hydrodynamic simulations of the \pup system with \textsc{AstroBEAR}.  We use the radiative transfer code \textsc{RADMC-3D} to construct the broad-band spectral-energy-distribution (SED) and synthetic observational images from our simulations.  Given the reported binary parameters, we are able to reproduce the current observational data if a short pulse of dense material is released from the AGB star with a velocity sufficient to escape the primary but not the binary.  Such a situation could result from a thermal pulse, be induced by a periastron passage of the secondary, or could be launched if the primary ingests a planet.
\end{abstract}

\begin{keywords}
method: numerical --- stars: AGB and post AGB --- ISM: structure --- stars: winds, outflows 
\end{keywords}

\section{Introduction}

Recent observations of the \pup system suggest that the Asymptotic Giant Branch (AGB), Mira-like variable may be in the early stages of transitioning to a planetary nebula (PN) \citep{ker2015,ker2015b}.  At a distance of 64 $pc$, \pup is one of the nearest and brightest AGB stars. It seems to be orbited by a close binary companion, and thus represents a unique laboratory in which to test models of the late-stages of stellar evolution.  

Adaptive optics imaging has revealed the presence of a optically thick, circumstellar disk with wide-bipolar outflows \citep{ker2015}.  Its M5III spectral type \citep{dumm1998} implies that \pup has an effective temperature $\sim 3500\ K$. Photometric studies show that it is a variable star with an apparent magnitude varying from 2.60 to 6.00 with a stable period of 140.6 days \citep{samus2009,bedding2002}, consistent with a typical period associated with thermal pulsation \citep{bowen1988}. The derived radius of \pup is $123\pm14R_{\odot}$ \citep{ker2014} thus its luminosity is $2000\pm700L_{\odot}$. Placing \pup on a Hertzsprung-Russell diagram using a ZAMS evolution model computed by \cite{bertelli2008} implies that \pup is now an AGB star with mass of $2^{+1.0}_{-0.5}M_{\odot}$ and an age of $1.5^{+1.5}_{-1.0}Gyr$ .

\pup has a large lobe structure that extends more than $10\ AU$ to the northeast of the disk in L-band images \citep{ker2014}, likely due to the interaction of an AGB wind with the secondary star. A recent result by \cite{ker2015} supports this hypothesis, revealing evidence of a close-in secondary source at a projected separation of $2\ AU$. In addition, the existence of an optically-thick circumstellar dusty disk hints at  the presence of a secondary. If the disk were stable and in approximate Keplerian motion, it would
have a high specific angular momentum. However, AGB stars are slow rotators \citep{meibom2009} and the large difference of specific angular momentum implies that there should be some mechanism that can transfer angular momentum to the gas and dust. A companion can transfer angular momentum to the gas and shape the outflows seen in the post-AGB and PN phases \citep{nordhaus2006,nordhaus2007}.

We have carried out a hydrodynamic simulation with \textsc{AstroBEAR} and have used the result to produce synthetic images from Monte Carlo radiative transfer with the code \textsc{RADMC-3D}. 
\cite{hendrix2015} used a similar method. We describe the hydrodynamic simulation in detail in Sect.~\ref{hydromodel} and present the results in Sect. \ref{hydroresults}.   We discuss the model used in our \textsc{RADMC-3D} simulation to generate  synthetic observations in Sect.~\ref{radmcmodel}.  In Section~\ref{rtresults} we compare our synthetic observations to the observational data.

\section{Hydrodynamic model description}\label{hydromodel}
We have performed three-dimensional hydrodynamic adaptive-mesh-refinement (AMR) simulations of a $2\ M_{\odot}$ AGB star with a $0.5\ M_{\odot}$ secondary at a separation of $2\ AU$\footnote{It was $4\ AU$ in the published version which was wrong}. The AGB star is represented by a sphere of radius $1\ AU$ with a steady mass-loss rate of $\dot{M}_s=1.65\times10^{-7}M_{\odot}/{\rm yr}$ and wind velocity of $v_s=39.3 km/s$. After the simulation reaches steady-state, we eject a pulse of dense wind at a rate of $\dot{M}_p=9.30\times10^{-6} M_{\odot}/{\rm yr}$ that lasts $11.6$ years. The speed of the pulsed material is $v_p=22.2$ ${\rm km/s}$ which is above the escape velocity of the AGB star but below the escape velocity of binary system.

Our chosen  value for mass loss rate is comparable to normal AGB winds but our initial wind velocity is higher. We choose a high velocity at the initial launch radius because we do not include wind launching in our models.  This means the gravitational force of the AGB star will only decelerate the wind as it expands outwards and we seek to keep $v_w$ in the $\sim 15\ km/s\sim20\ km/s$ range at radii of interest where the interaction with the binary occurs.  A model for driving AGB winds \citep{bowen1988,feuchtinger1993} will be included in our future work.

The properties of such a pulse are consistent with those expected from the ingestion of a planet by the AGB star. Such events are expected to occur on the giant branches if planetary companions are present and initially orbiting within $\sim10\ AU$ of the main-sequence progenitor \citep{nordhaus2010,nordhaus2013}. Radiation from the AGB star is assumed to be isotropic at all radii with the dust and gas fully coupled.  Under such conditions, the radiation force can be expressed as $f_{rad}=\rho\kappa_{total} L/c r^2$ where $\kappa_{total}$ is the mass weighted opacity of gas and dust. The radiation force is proportional to the gravitational force $f_{grav}=G M \rho/r^2$ in the optically thin limit \citep{chen2016}. The magnitude of the total inward force acting on the gas is therefore,
\begin{equation}
f=f_{grav}-f_{rad}=\alpha f_{grav}
\label{eq:1}
\end{equation}
We assume that $\alpha=0.134$ in our model.  Note that once the disk forms, radiation may not be isotropic at all radii outside of the AGB star. The formation of an optically thick region, such as a torus or disk, would block photons and the local interior regions of these structures would experience a reduced or absent direct radiation force.

Optically thin regions, such as those in the polar direction (and for the majority of our computational domain) would  experience the full radiation force.  Despite our use of Eq.~\ref{eq:1} everywhere, a prominent disk clearly forms suggesting that any case with a position dependednt reduced radiation force (less total outward force)  would also incur disk formation. Future work should incorporate a more accurate local variation in the radiation force.

Assuming that the dust and gas are in local thermal equilibrium (LTE) and that the luminosity from the AGB star is constant, the gas temperature varies as $T\sim\sqrt{r}$.  At $r\sim$ $120\ R_{\odot}$ the surface temperature of the AGB star is $3500\ K$ and is approximately $400\ K$ at $40\ AU$ away.  Since our simulation zone is $(80\ AU)^3$ we employ an isothermal $T=400\ K$ for  our simulation. We use isothermal condition for practical purpose and the actual temperature in the disk might be close to $400\ K$. We know that the temperature near the actual AGB star is much higher than $400\ K$ and that temperature gradient is an important condition for driving AGB winds, but these effects are are only important in the launch region which we do not model.  As noted above, we leave an exploration of the effect detailed driving mechanism of AGB winds and cooling for future research.

To carry out our three-dimensional hydrodynamic simulations, we employ \textsc{AstroBEAR}. \textsc{AstroBEAR} is a multi-physics, adaptive-mesh-refinement (AMR) code which employs a Riemann solver to solve the fluid equations \citep{carrol2013}. In our simulation, we use $100^3$ computational cells as the base grid and 3 levels of AMR such that the total effective resolution is $800^3$. The equations of motion of fluid in the simulation are
\begin{equation}
\frac{\partial\rho}{\partial t}+\triangledown\cdot\left(\rho\mathbf{v}\right)=0
\end{equation}
\begin{equation}
\frac{\partial\rho\mathbf{v}}{\partial t}+\triangledown\cdot\left(\rho\mathbf{vv}\right)=-\triangledown p-\frac{\alpha G M_1\mathbf{\hat{r_1}}}{r^2_1}-\frac{G M_2 \mathbf{\hat{r_2}}}{r^2_2}
\end{equation}
\begin{equation}
p=n k_b T,
\label{eq:2}
\end{equation}
where subscript $1$ represents the AGB star and subscript $2$ represents the secondary. The mean atomic weight of the gas material is $1.3$ $m_H$.

\section{Disk and outflow formation}\label{hydroresults}
We run the simulation for $1315$ years. Approximately, $110$ years after the ejection of the dense wind, material falls back with a disk forming in the orbital plane. Meanwhile, low-density outflows form in the polar directions. In Figure 1 and 2 we show results of the simulations in terms of cuts of density in, and perpendicular, to the equatorial plane (Fig 1) as well as 3-D iso-density contours (Fig 2).

To measure whether material is gravitationally bound to the binary system, we construct the parameter 
\begin{equation}
q\equiv\frac{e_k + e_{interal}}{\|\phi\|}=\frac{0.5 v^2+1.5 k_b T}{\|\phi\|}
\end{equation}
where $\phi$ is the specific gravitational potential energy of the material.  When $q<1$ the material is bound to the system.

\begin{figure}
\centering
\includegraphics[width=8.5cm]{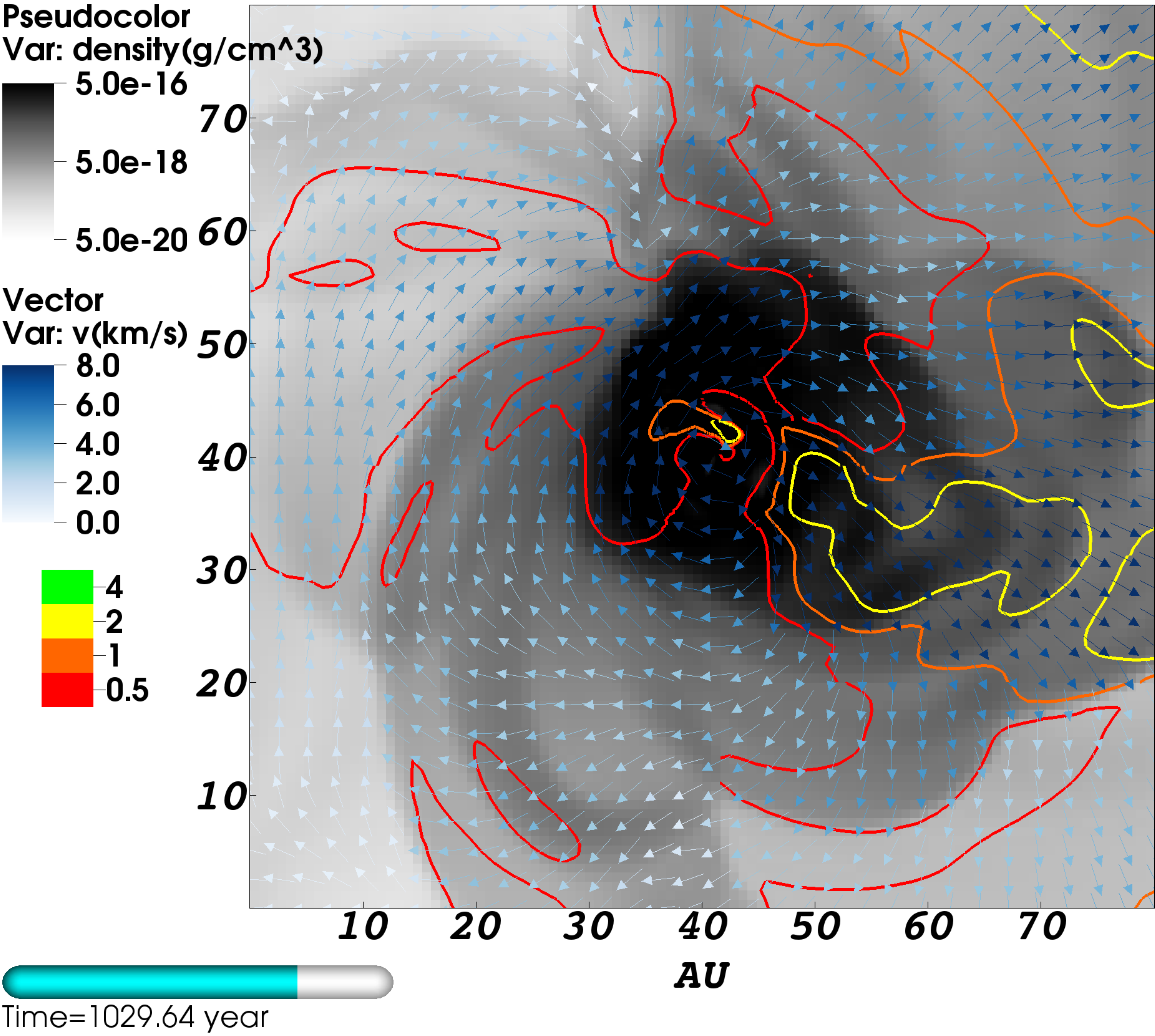}
\includegraphics[width=8.5cm]{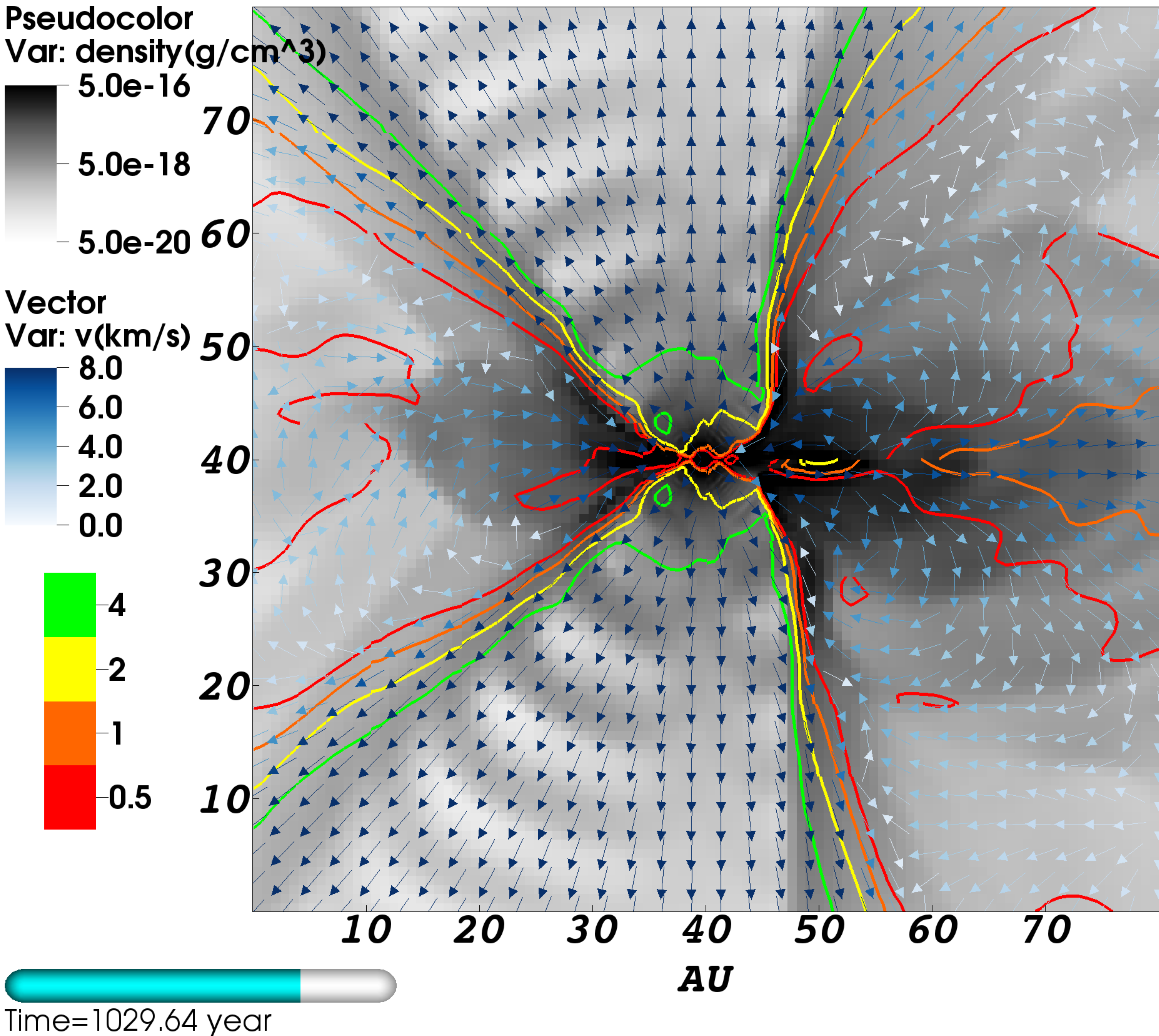}
\caption{Top: Face-on view of the disk (x-y plane).  Bottom: Edge on view of the disk (x-z plane). Density values are shown with a gray-color map, while velocity vectors are shown in white and blue at the end of the simulation.  Four contours of $q$ delineate the bound and unbound regions.}
\label{fig:1}
\end{figure}

\begin{figure}
    \centering
    \includegraphics[width=8.5cm]{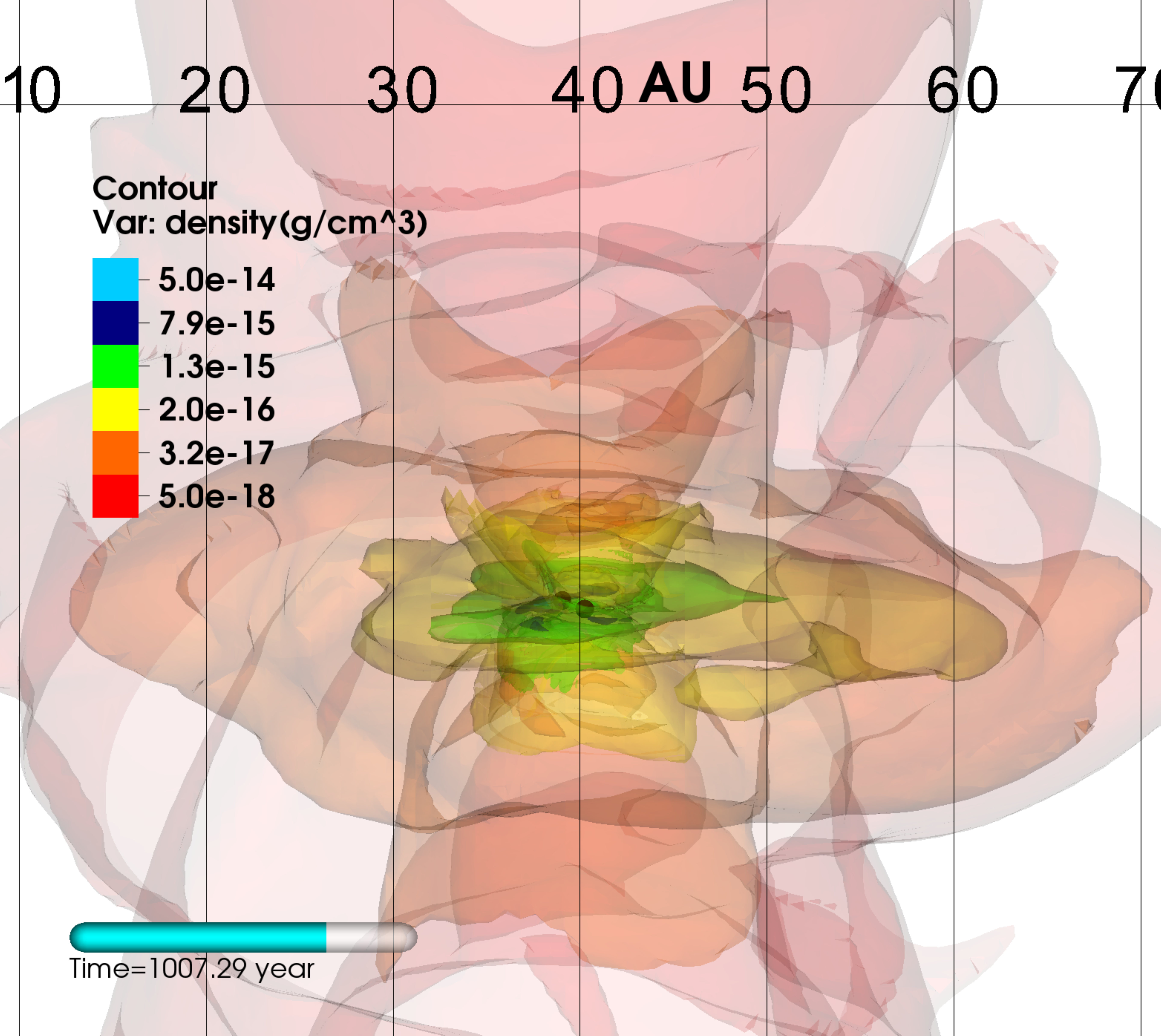}
    \caption{3D contour plot of density. The sight inclination is $73.3^{\circ}.$}
    \label{fig:contour}
\end{figure}

The density cuts in figure~\ref{fig:1} include conturs of $q$ and they demonstrate
the formation of both unbound bipolar outflows and a gravitationally bound disk structure. In the top panel, we see high-density gas orbiting the binary in a torus/disk configuration. The radius of the disk is roughly $12\ AU$ and there is a spiral-arm in its outer regions which sweeps clockwise. The $q=0.5$ and $q=1$ iso-contours demonstrate that most of the gas in mid-plane is gravitationally bound, although there is some escaping gas in the lower right corner of the top panel. This is to be expected as gas from the mass-losing AGB star continually interacts with the disk. The disk should be sustible to Rossby wave instabilities \citep{meheut2012} which may be the cause of the episodic ejection.  

In the bottom panel, the $q=4$ the contours take the form of bipolar cones indicating that the gas in the polar direction is not bound. Note the presence of "ripples" in the two cones induced by the orbital motion of the binary system. The velocity of the bipolar outflows is roughly $20\ km/s$. Note that the disk is  seen in the bottom panel of fig.~\ref{fig:1} as the dense gravitationally-bound region inside the red contour. We note that in the equatorial region, some of the outflowing gas will fall back onto the the mid-plane and incorporate into the disk while some will escape the system.

Figure~\ref{fig:contour} shows the 3D iso-density plot with the disk spin axis inclined by $73.3^{\circ}$ to the direction toward the observer. 
The bipolar lobes are seen in the red and yellow iso-density contours in this image. The green regions demarcate a iso-density contour of $1.3\times10^{-15}\ g/cm^3$ with includes the disk. 

\section{\textsc{RADMC-3D} model}\label{radmcmodel}
We employ the \textsc{RADMC-3D}	code \citep{dullemond2012} to post-process our hydrodynamic results and create synthetic maps for comparison with observations. Given a prescribed dust distribution, we use a Monte Carlo Method \citep{bjorkman2001} to determine the local dust temperature, which is then used to produce images of photon density with asymmetric scattering. The results of our hydrodynamic simulation serve as the input for our radiation transfer simulation.

\subsection{Spectral Energy Distribution}
We adopt an AGB photospheric model from \cite{castelli2004} which is shown as the blue curve in Figure~\ref{fig:SED}. The AGB model has an effective temperature of $3500\ K$, $\log{g}=1.5$ and $\left[ M/H\right] =0.0$. The raw SED is subject to reddening of $E\left( B-V \right)=0.6$ with a standard Milky Way $R_V =A_V /E\left( B-V \right) = 3.1$ interstellar dust model and shown as the green curve in Figure~\ref{fig:SED} \cite{fitz1999}. 

\subsection{Dust species and spatial distribution}
Observational studies of \pup show that there is an infrared excess in the SED.  \citep{ker2014,ker2015}. Such infrared excesses in evolved stars are associated with dust shells or disks \cite{nordhaus2008}.  Furthermore, the $10$ $\mu$m feature suggests the dust is primarily silicate-based.  As such, we choose two kinds of amorphous silicates: \ce{MgFeSiO_4} olivine  and $MgFeSi_2 O_3$ pyroxene \citep{jaeger1994,dorschner2004} which are the most common silicates around AGB stars.  We assume that both types of dust grains dust are spherical and with size- number distributions of $dn\sim a^{-3.5}da$ \citep{mathis1977},  where $a$ is the dust radius and $dn$ is the number density. We consider dust radii in the range from $0.1\mu m\le a \le 0.3\mu m$. We choose this range of radii because:  1. the flat feature in SED between $1$ $\mu m$ to $4$ $\mu m$ implies that the disk may have dust grains with radii around $0.5$ $\mu m$. 2.  larger dust grains need more time to evolve from small dust grains 
and large  grains may not obey the size distribution law described in \citep{mathis1977}. We thus set the cutoff at $0.3$ $\mu m$.

Dust cannot form when the environment is too hot. If we assume the dust to be in approximate LTE with the star's effective temperature, given the luminosity $L=2000\pm700L_{\odot}$ and the dust sublimation temperature $T_{sub}=1500\ K$, the dust is unlikely for form within $3.2\ AU$. However, there are likely high temperature shocks around the star created by the AGB winds and the orbital motion of the secondary. The high temperature shocks will moderately extend the dust free region. As such, we set the dust-to-gas mass ratio $\eta=0$ within $4\ AU$ of the center-of-mass of the binary.   For the rest of the computational domain, we set $\eta=0.01$, since small dust grains are almost fully coupled with the fluid \citep{mastrodemos1998}. We assume that $80\%$ of the dust mass  is olivine and $20\%$ mass of the dust is pyroxene. The total mass of the dust in the simulation is $8.55\times10^{-8}M_{\odot}$.

\begin{figure}
\centering
\includegraphics[width=7.5cm]{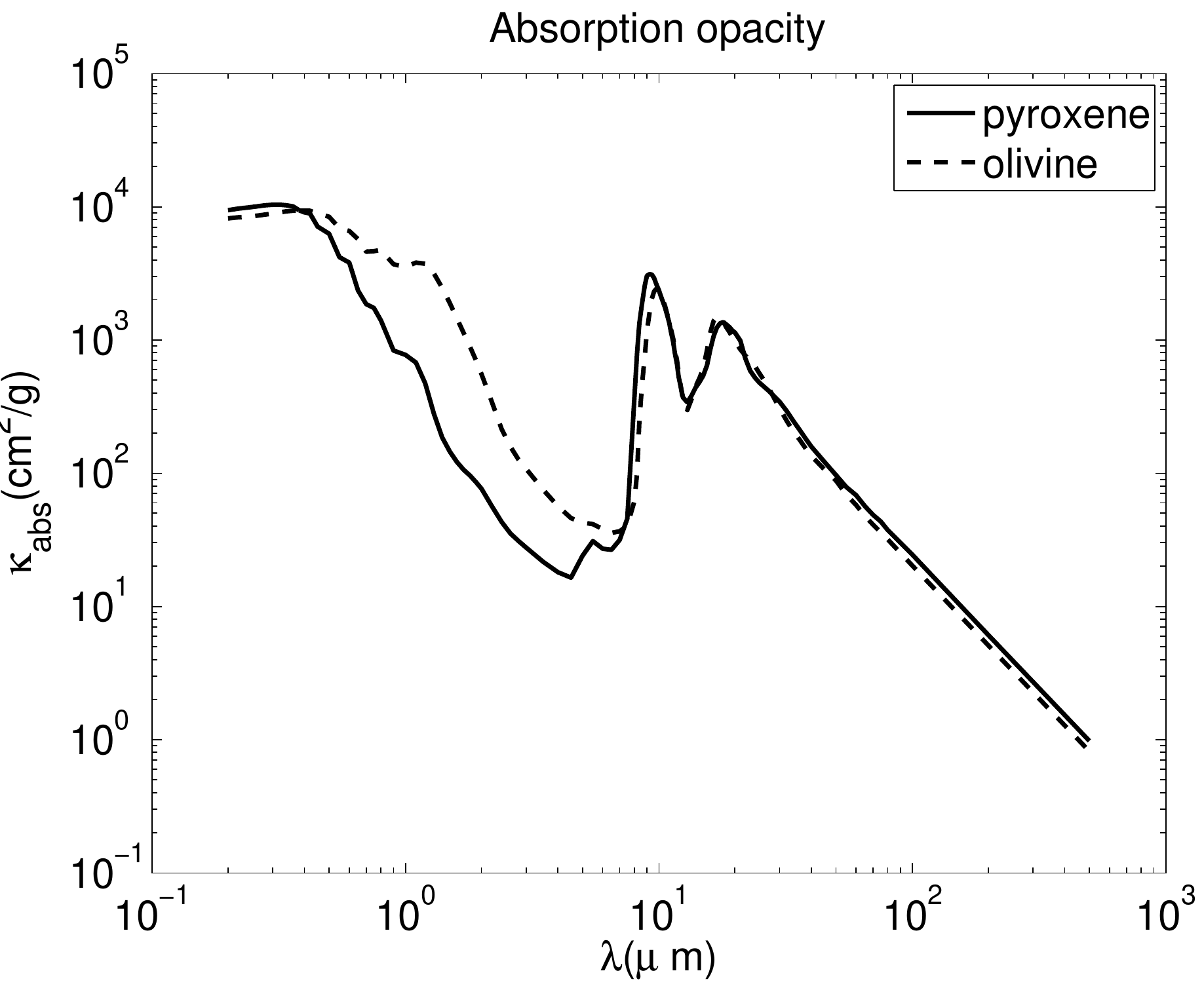}
\includegraphics[width=7.5cm]{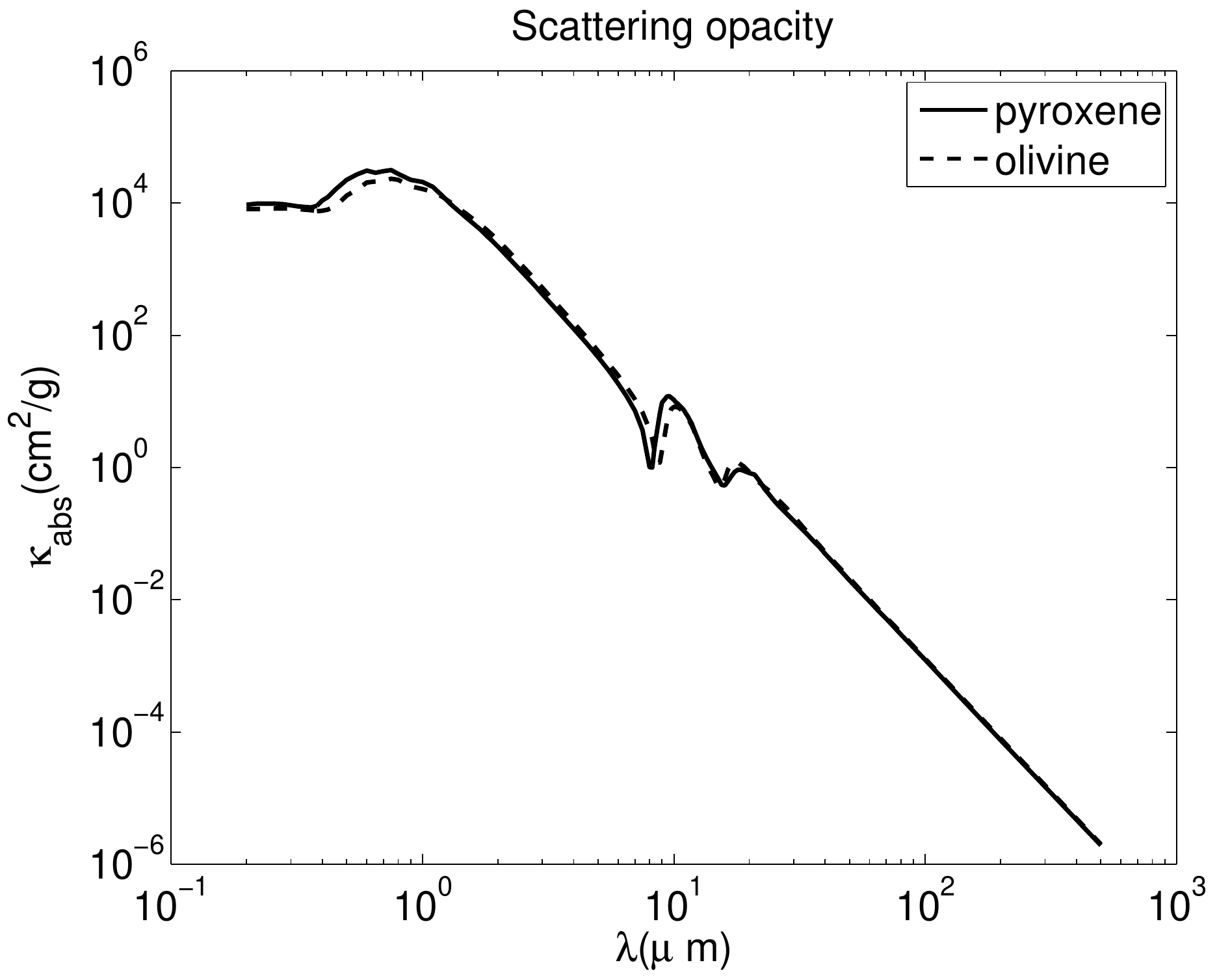}
\caption{The wavelength dependent absorption and scattering opacity of the dust mixture.}
\label{fig:dust}
\end{figure}

Given our chosen dust size distribution, dust species and optical properties of amorphous silicates, the optical properties of the dust mixture can be computed from Mie theory \citep{vandehulst1957}. We use the MATLAB code \citep{matzler2002} to calculate the absorption and scattering opacities and the results are plotted out in figure ~\ref{fig:dust}.  We also take asymmetric scattering by dust into consideration. In our Monte Carlo simulation, we use the Henyey-Greenstein \citep{henyey1941} scattering phase function 
\begin{equation}
    p_{\lambda}(\theta)=\frac{1}{4\pi}\frac{1-g_{\lambda}^2}{(1+g_{\lambda}^2-2g_{\lambda}\cos(\theta))^{3/2}}
\end{equation}
where $-1\leq g_{\lambda} \leq 1$ is the wavelength dependent asymmetric coefficient. The actual asymmetric coefficient for dust is close to $1$ in short wavelength regime. When the asymmetric coefficient is close to $1$, the scattering is strongly forward directed.

In reality, dust tend to agglomerate into larger grains when the temperature drops \citep{gail2013} and larger grains are more weakly coupled to the fluid \citep{vanmarle2011}. Therefore,  by assuming a fixed dust distribution, composition, and gas-to-dust ratio,
our treatment may underestimate the dust contribution in dense regions such as the disk itself \citep{woitke2006}. However, this simplifying  approach for present purposes makes the problem tractable with current computational resources.  We leave the self-consistent, multi-fluid, three-dimensional, radiation-hydrodynamic modeling for future work.  

\section{Radiation transfer simulation results}\label{rtresults}

\begin{figure*}
\centering
\includegraphics[width=18cm]{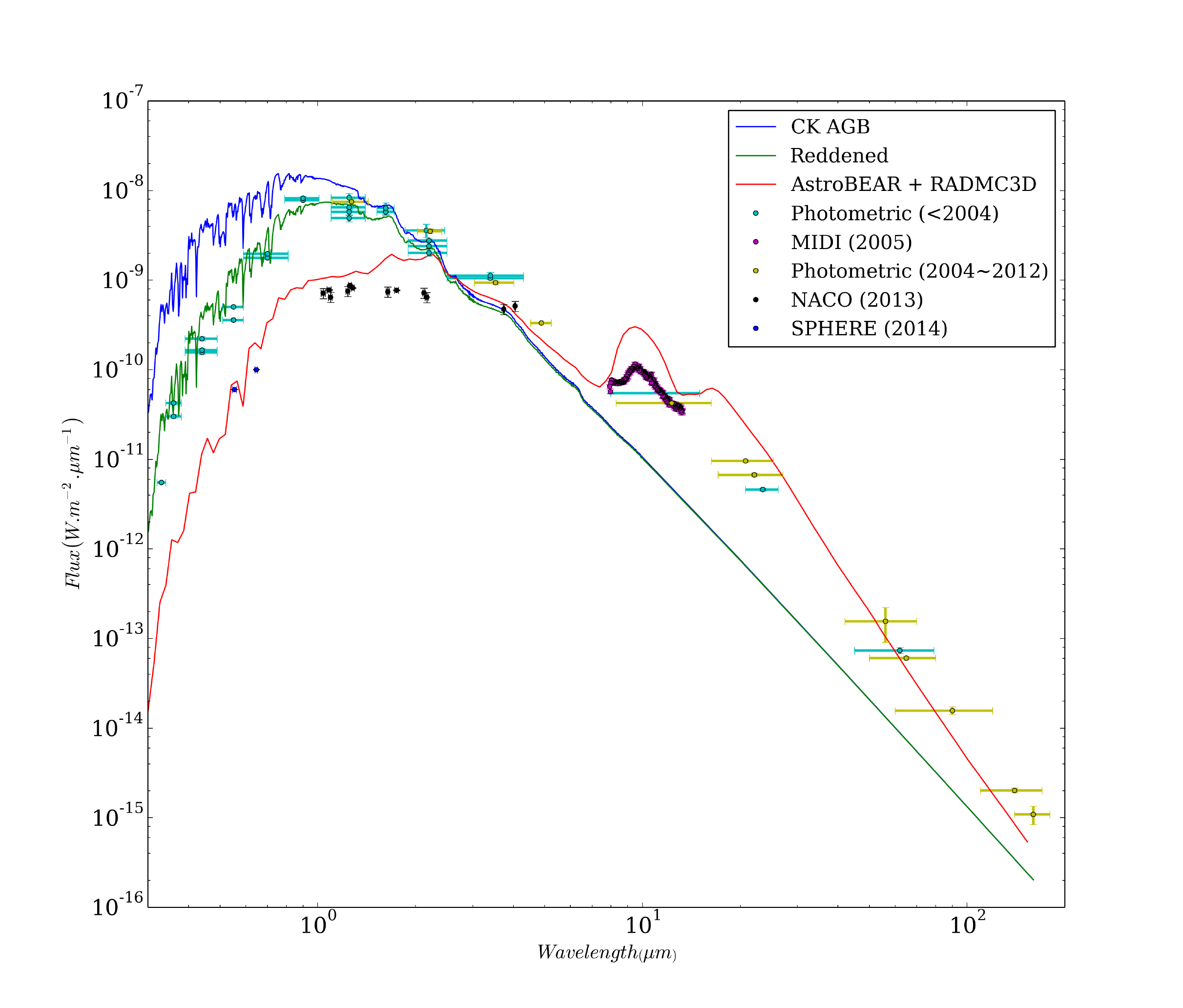}
\caption{Spectral energy distribution of \pup.  Current optical flux measurements show an order-of-magnitude decrease over the last decade.}
\label{fig:SED}
\end{figure*}

Figure~\ref{fig:SED} shows our results for the broad-band SED computed from our hydrodynamic model and compared with existing observational data (details can be found in \cite{ker2014,ker2015}). In Figure 5 we also present synthetic observations in terms of N and V band images of models and compare these with recent ZIMPOL observations.

Consideration of fig ~\ref{fig:SED} and Figure 5 shows we successfully reproduce the $1 \mu m$ to $4 \mu m$ flat features as well as the $10 \mu m$ bump in the SED.  We note that the NACO data, indicating the flat, lower intensity regions of the spectrum between $1 \mu m$ to $4 \mu m$ were taken in 2013 and represents higher quality data than the photometeric data associated with the pre-2004 points.  The \textsc{AstroBEAR} simulation, with our choice of dust model, thus is able to capture the shape of the SED across 2 orders of magnitude in radius.

\begin{figure}
\centering
\includegraphics[width=4.1cm]{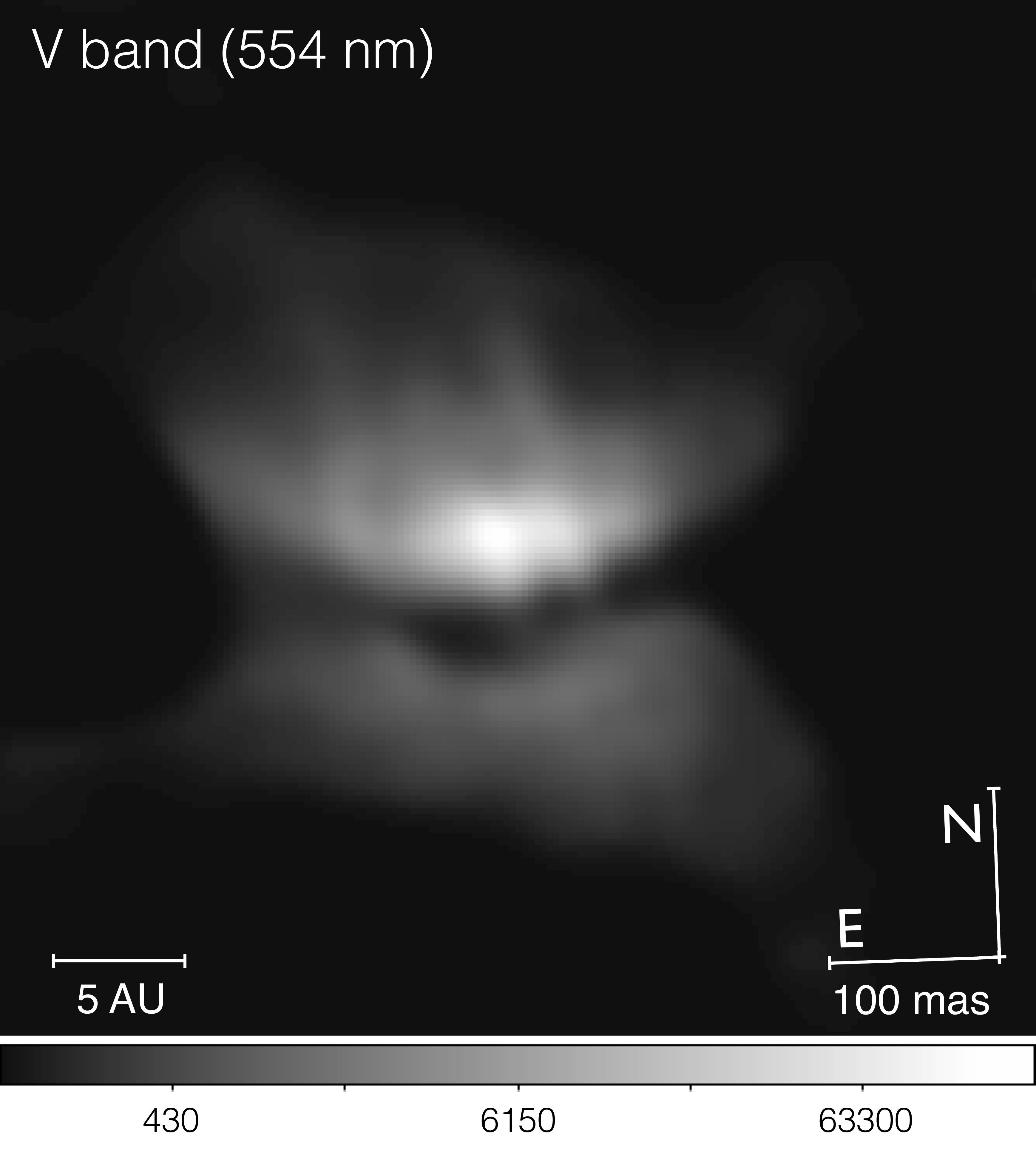}
\includegraphics[width=4.1cm]{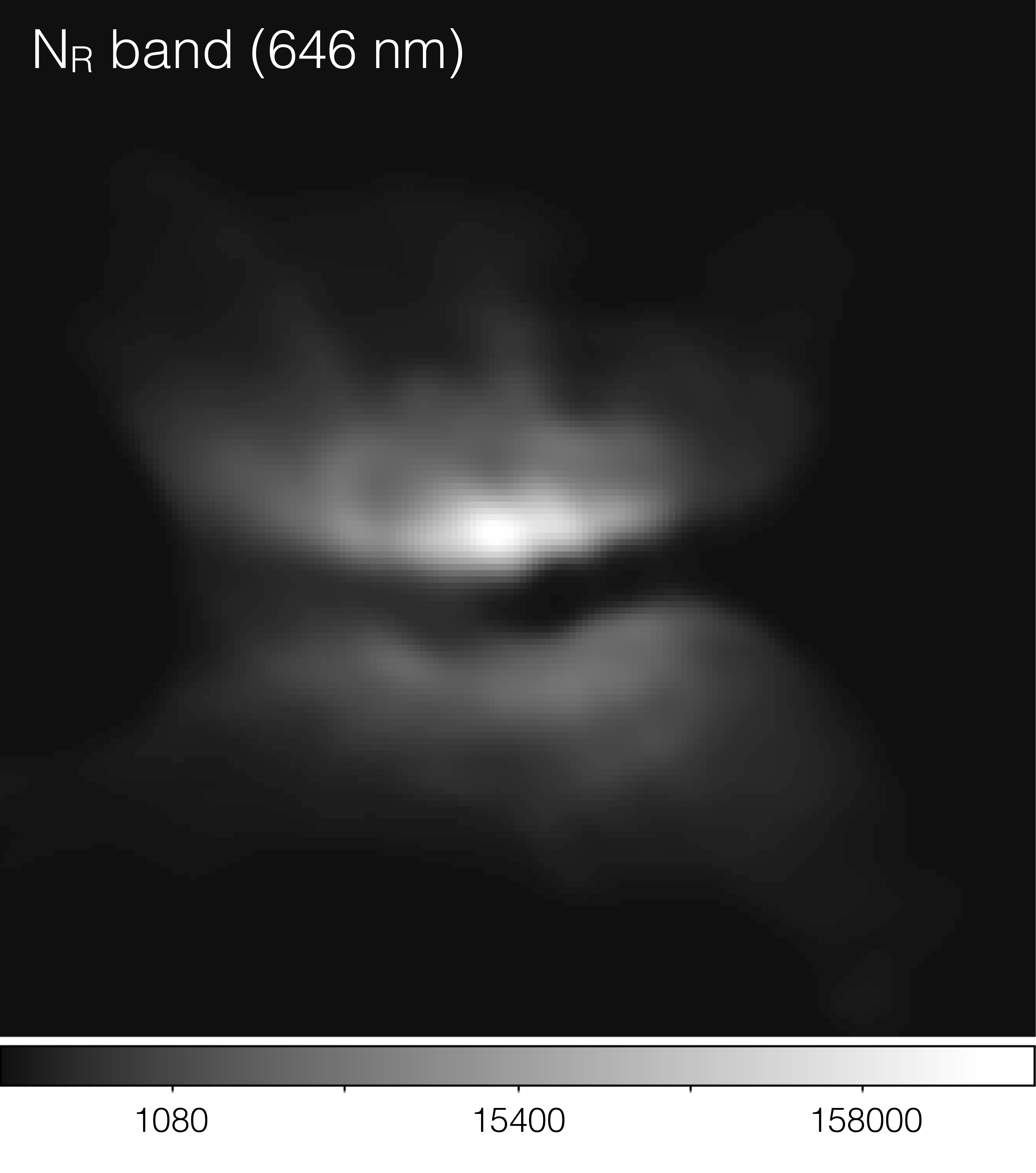}
\includegraphics[width=7.5cm]{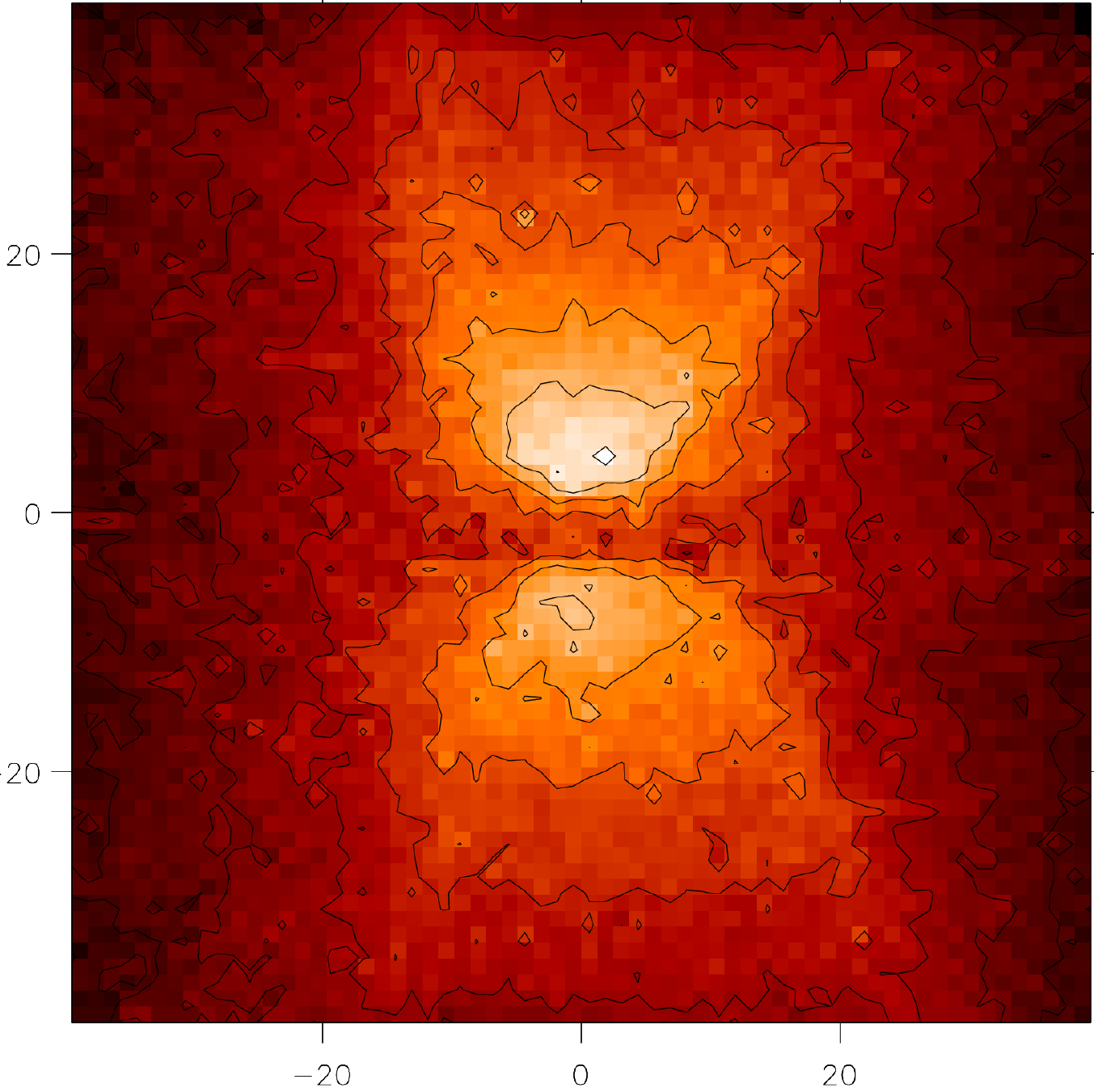}
\includegraphics[width=7.5cm]{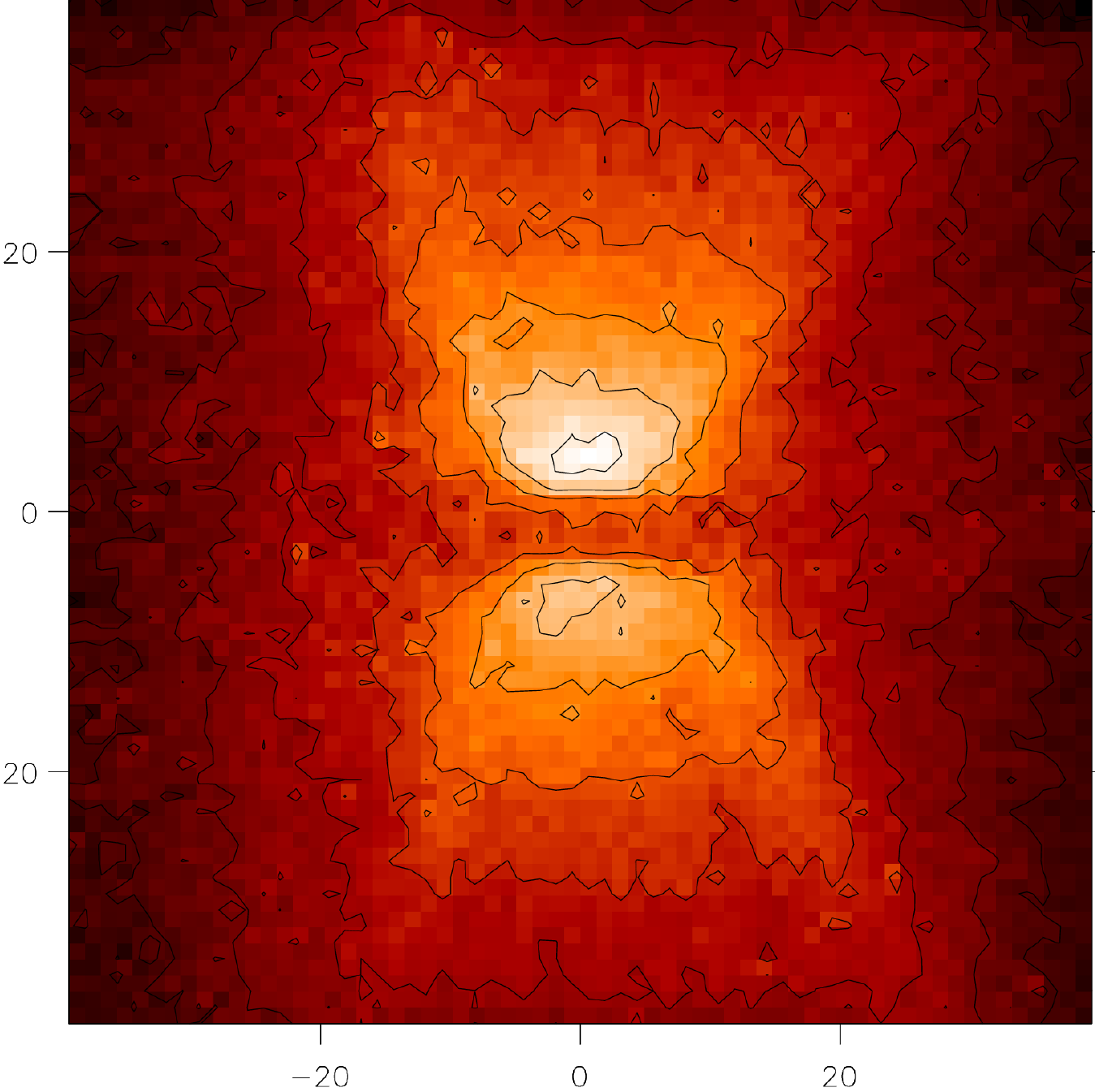}
\caption{Synthetic V-band and N-band images from our hydrodynamic simulation compared to the corresponding ZIMPOL observations for the same logarithmic scale. The synthetic image in the middle is the V-band and the bottom image is the N-band. The dimension and the inclination of our images are $40\ AU\times40\ AU$ and $82^{\circ}$ respectively.}
\label{fig:vnbandside}
\end{figure}

In Figure~\ref{fig:vnbandside}, we have compared synthetic V-band and N-band \textsc{RADMC-3D} images with the corresponding ZIMPOL observations \citep{ker2015}. The ZIMPOL observations have an angular resolution of $20\ mas$, which corresponds to $1.28\ AU$ in \pup. To foster the best comparison, we made the resolution in the synthetic image the same as that of the observation. The images reveals an optically thick circumstellar torus in V and N band.

\begin{figure}
\centering
\includegraphics[width=4.15cm]{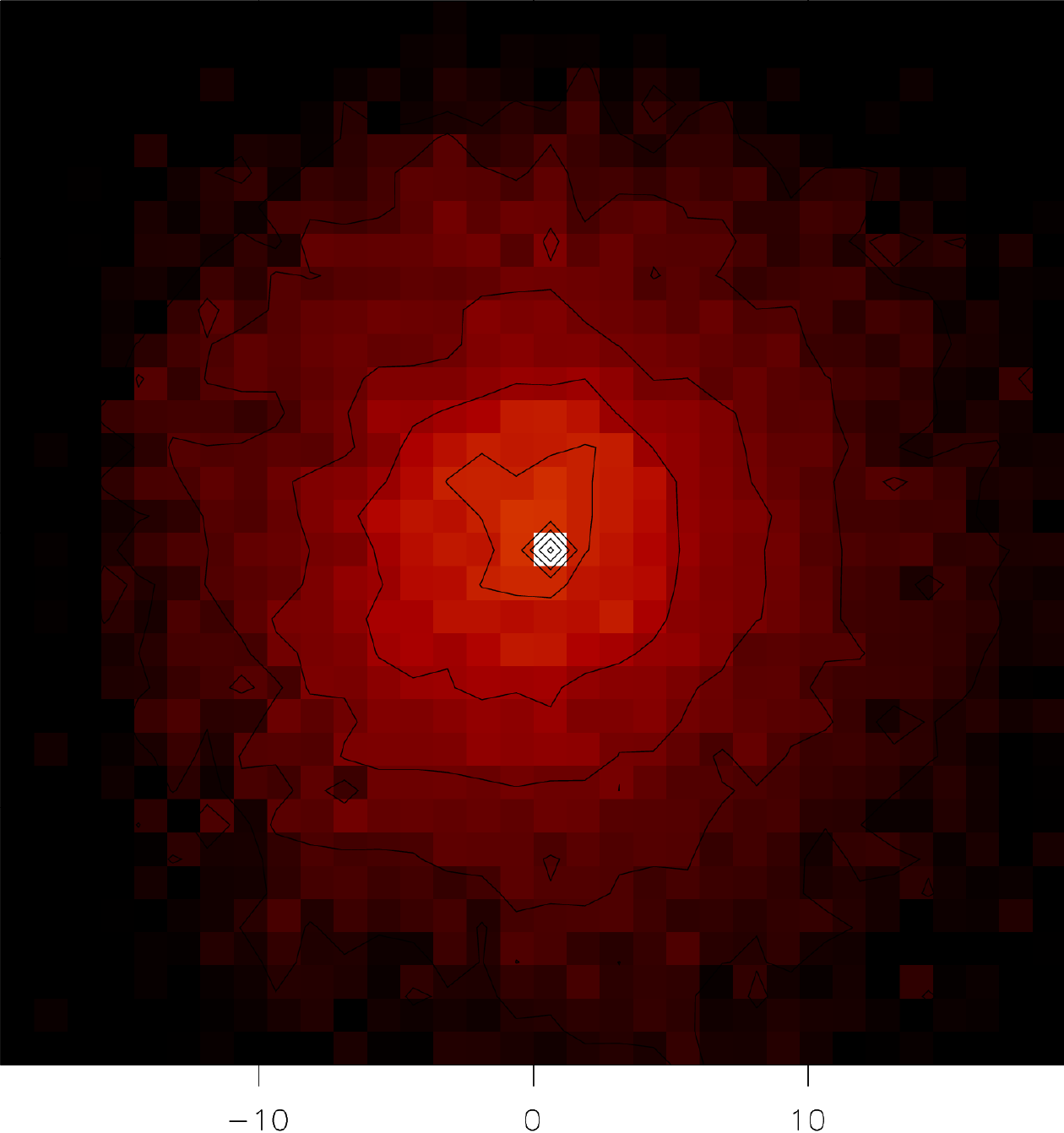}
\includegraphics[width=4.15cm]{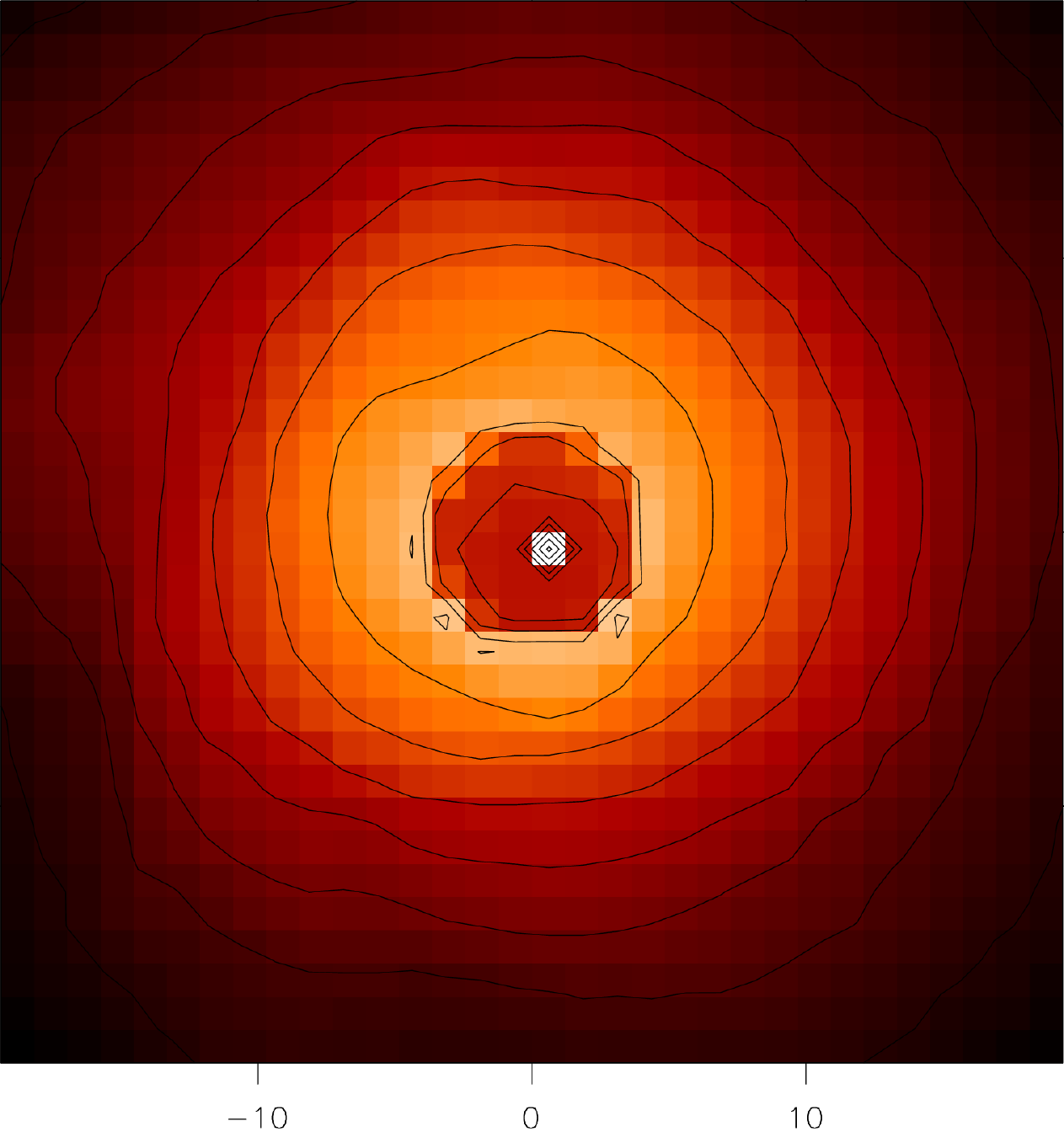}
\caption{The image on the left shows the top-down synthetic image of $600\ nm$ from top. The image on the right shows the synthetic image of $11\ \mu m$ also from top.}
\label{fig:vtop}
\end{figure}

The fit between our simulation and observations is good as the simulation synthetic images show both the bipolar lobes and dark band indicating the presence of the disk.  The broad opening angle of the lobes is also recovered in the simulation. Finally in figure~\ref{fig:vtop}, the $600\ nm$ synthetic image shows a brighter center while the $11\ \mu m$ synthetic image has a brighter ring. The $11\ \mu m$ photon is a signature of silicate dusts, thus the $11\ \mu m$ synthetic image shows where the dusts are. The contour lines show that the inner boundary of the ring is about $4\ AU$ which is the no-dust radii we have assumed. The relatively dim center implies that there are very few dusts in the polar direction. It tells us that the optical depth in the polar direction is much lower compared to the equatorial plane - otherwise we would have seen a brighter center. Given that there are many more dusts outside the center, the brighter center in the $600\ nm$ image indicates that the dusts in the ring are colder - otherwise they will emit more short wavelength photons. It also proves that the polar direction is optically thin.

\section{Summary and discussion}
In this paper, we present a fully 3-D hydrodynamic simulation that models a pulse of dense wind followed by constant stellar wind in a binary system.  Our simulation self-consistently forms a circumbinary disk with wide-bipolar outflows for the \pup system parameters. Using the output of  hydrodynamic modeling to produce synthetic observations of broad-band photometric and imaging data significantly improves upon previous non-dynamical, morphological studies of \pup.  

Throughout our computational domain, we tracked the fluid to determine  whether it is gravitationally bound or escaping.  The bulk of the bound material is located in the circumstellar disk in the orbital plane.  Along the poles, a wide angle low velocity outflows emerges with measured velocities of $\sim20\ km/s$.  

Synthetic observations constructed by post-processing our hydrodynamic simulation with the radiation transfer code \textsc{RADMC-3D} shows strong morphological similarity to the V-band and N-band SPHERE observations. The broad-band SED computed from our results also matches the observations well, reproducing the "flat" range from  $1 \mu m$ to $4 \mu m$ and the infrared excess seen long-ward of $10 \mu$m. The top-down synthetic images show that the polar direction is optically thin and the dust in the disk has low temperature.

This study is an initial step in fully dynamical three-dimensional simulations of evolved binary star systems and their photometric and imaging observations. Future work should include additional physical processes such as cooling, self-consistent radiative transfer and multi-dust species.

\section{Acknowledgements}

We gratefully acknowledge the help from Dr. Eric Lagadec for the useful discussion of the recent observations of \pup that led to this paper. We also thank Dr. Pierre Kervella and Dr. Miguel Montarges very much for sharing us the useful data and images. Our thanks should also be given to Dr. Baowei Liu and Dr. Jonathan Carroll-Nellenback for helping us in developing the code. This work used the computational and visualization resources in the Center for Integrated Research Computing (CIRC) at the University of Rochester. Part of the work also used the Extreme Science and Engineering Discovery Environment (XSEDE), which is supported by National Science Foundation grant OCI-1053575. Financial Support for this project was provided by the Department of Energy grant GR523126, the National Science Foundation grant GR506177, and the Space Telescope Science Institute grant GR528562.

\bsp

\label{lastpage}

\end{document}